\documentclass[12pt]{article}

\usepackage{mathrsfs}
\usepackage{amssymb}

\usepackage{graphicx}
\usepackage{epsf}
\usepackage{graphics}
\usepackage{psfrag}

\usepackage{cite}
\newcommand{\be}{\begin{equation}}
\newcommand{\ee}{\end{equation}}
\newcommand{\bea}{\begin{eqnarray}}
\newcommand{\eea}{\end{eqnarray}}

\newcommand{\rmd}{\mathrm{d}}

\def \ch{{\cal H}} 
\def \dc{\delta_c}
\def \dx{\delta_x}
\def \cshat{\hat{c}_s^2}
\def \cshatn{\hat{c}_s}

\def \fac{{\cal F}}

\long\def\symbolfootnote[#1]#2{\begingroup%
\def\thefootnote{\fnsymbol{footnote}}\footnote[#1]{#2}\endgroup}

\begin{document}

\setcounter{footnote}{0}
\setcounter{figure}{0}
\setcounter{table}{0}

\vspace{-0.5cm}
\thispagestyle{empty}
\begin{flushright}
{\small CERN-PH-TH/2008-160}\\
{\small IFT-UAM/CSIC-08-46}
 \end{flushright}

\vspace{1.0cm}

\begin{center}
{\Large\textbf{Parameterizing the Effect of Dark Energy}} 
\vskip 0.1cm
{\Large\textbf{  
Perturbations on the 
Growth of Structures}} 
\vspace{1.1cm}

\textbf{Guillermo Ballesteros}\footnote{guillermo.ballesteros@uam.es}  \\ \vspace{0.3cm}
 {\it Instituto de F\'isica Te\'orica UAM/CSIC, Universidad Aut\'{o}noma de Madrid,
Cantoblanco, 28049 Madrid, Spain}\\ \vspace{0.1cm}
 \textit{CERN, Theory Division, CH-1211 Geneva 23, Switzerland}\\
\vspace{0.42cm}
and \\
\vspace{0.42cm}
\textbf{Antonio Riotto}\footnote{antonio.riotto@cern.ch} \\ \vspace{0.3cm}
\textit{INFN, Sezione de Padova, Via Marzolo 8, I-35131, Padova, Italy}\\ \vspace{0.1cm}
\textit{CERN, Theory Division, CH-1211 Geneva 23, Switzerland}
\end{center}

\vspace{0.5cm}

\begin{center}
{\bf Abstract}
\end{center}
\vspace{-0.2cm}
We present an analytical fit to 
the growth function of the dark matter perturbations 
when dark energy perturbations
are present. The growth index $\gamma$ depends upon the dark energy equation of state $w$, the speed
of sound of the dark energy fluctuations, the dark matter abundance and the observed comoving scale. The growth index changes by \mbox{${\cal O}(5\%)$}
for small speed of sound and large deviations of $w$ from $-1$ with respect
to its value in the limit of no dark energy perturbations.

\newpage
\setcounter{page}{1}

\section{Introduction}
Current observations of Type Ia supernovae luminosity distances indicate 
that our Universe is in a phase of accelerated expansion
\cite{Astier:2005qq}. 
Various proposals have been put forward to explain the present acceleration of the Universe.  One  can roughly distinguish two classes. 
On  the one hand, the acceleration might be caused by  the presence of  dark energy, a fluid with negative equation of state $w$. This may be provided 
by a tiny  cosmological constant which is characterized by \mbox{$w=-1$} or by  some ultralight scalar field whose potential is presently 
dominating the energy density of the Universe. This is usually dubbed quintessence \cite{ArmendarizPicon:2000dh} (see \cite{Copeland:2006wr} for a comprehensive review). 
On the other hand, the acceleration might be due to 
a modification of standard gravity at large distances. This happens in $f(R)$ theories \cite{Carroll:2003wy} and in extra-dimension inspired models, like DGP \cite{Dvali:2000hr}. Understanding which class of models Nature has chosen will represent not only a  breakthrough in cosmology, but also 
in the field of high energy physics.  

Mapping the expansion of cosmic scales and the growth of large scale structure in tandem can provide insights to distinguish between the two possible
origins of the present acceleration. For such reason, there has been increasing interest in analysing the time evolution of the dark matter perturbation. 
Several recent works deal with characterizing the growth of dark matter perturbations in different frameworks 
  \cite{Knox:2005rg, Koyama:2005kd, Koyama:2006ef, Kunz:2006ca,   Uzan:2006mf, Carroll:2006jn, Bertschinger:2006aw, Linder:2007hg, Tsujikawa:2007gd, Acquaviva:2007mm, Jain:2007yk, Wei:2008vw, Wei:2008ig, Zhang:2008kx, Bertschinger:2008zb}. 

The evolution of the growth function of dark matter perturbations
\mbox{$g=\delta_c/a\,$,} which is the ratio 
between the perturbation $\delta_c$ and the scale factor of the Universe $a$, can be parameterized in a useful way using the growth index $\gamma$ \cite{Linder:2005in}, defined in Eq. \ref{gammadef}. In a pure matter-dominated Universe, $g$ does not evolve in time (remains equal to one) and $\gamma$ is zero. However, in the presence of a dark energy background, $g$ changes in time, $\gamma$ is different from zero and its value can be approximated by 
\be
\label{eqlind}
\gamma=0.55+0.05\left[1+w(z=1)\right],  
\ee
which provides a fit to the evolution of $g$ to better than 0.2\% for $-1\lesssim w$ and a broad range of initial conditions 
for the dark matter abundance \cite{Linder:2005in}. Typically, the growth index in modified gravity models turns out to be  significantly different (for instance $\gamma\simeq 0.68$ for DGP \cite{Linder:2005in}) and therefore it is in principle
 distinguishable from the one predicted for dark energy models\symbolfootnote[2]{However, see \cite{Kunz:2006ca} and \cite{Bertschinger:2008zb}.}. The available data on the growth of structures are still poor and there is a long way to go 
before we can talk about precision cosmology in this respect. The methods developed to study the growth of structure involve baryon acoustic oscillations, weak lensing, observations of \mbox{X-ray} luminous clusters, large scale galaxy surveys, {\mbox Lyman-$\alpha$} power spectra and the integrated \mbox{Sachs-Wolfe effect} on the Cosmic Microwave Background.  
There are however various works that use these kind of techniques to place constraints on the growth index 
(and some also on the equation of state of dark energy) as well as  forecasts for its determination based on future observations \cite{Huterer:2006mva, Amendola:2007rr, Sapone:2007jn, Nesseris:2007pa, Dore:2007jh, Mantz:2007qh, DiPorto:2007ym, Yamamoto:2007gd,  Heavens:2007ka, Acquaviva:2008qp}. In particular, it is found in \cite{Heavens:2007ka} using Bayesian methods that a next generation weak lensing survey like DUNE \cite{Refregier:2008js} can strongly distinguish between two values of $\gamma$ that differ by approximately $0.05\,$. 
The authors of \cite{Amendola:2007rr} made a forecast for the same kind of satellite proposal and concluded that it will be 
possible to measure the growth index with an absolute error of about 0.04 at 68\% confidence level. In \cite{Sapone:2007jn} a 
slightly bigger error of 0.06 at the same confidence level is given for a forecast based on baryon acoustic oscillations. 
Finally, for a combination of weak lensing, supernovae and Cosmic Microwave Background data an error of about 0.04\, is estimated 
in \cite{Huterer:2006mva} after marginalizing over the other cosmological parameters. 
Since the growth index  is approximately equal to 0.55, the nearest future observations should be able to 
determine it with a relative error of around 8\%.

While much effort has been put into determining the value of the growth index in dark energy and in modified gravity models,  
less attention has been devoted to the possible effect on $\gamma$ of  non--vanishing dark energy perturbations. 
The latter do not affect the
background evolution,  but are fundamental in determining the dark energy
clustering properties. They will have an effect on
the evolution of fluctuations in the matter distribution and, consequently, on $\gamma$. While minimally coupled scalar field (quintessence) models commonly have a
non-adiabatic speed of sound close or equal to unity, and therefore dark energy perturbations can be neglected for them;
 other non-minimal models, for instance the adiabatic Chaplygin gas model, motivated by a
rolling tachyon \cite{Gibbons:2002md},  have  a speed of sound which is 
approximately zero. 
Observational implications of dark energy perturbations with a small  speed of
sound in a variety of dark energy models have been  recently
discussed in  k-essence \cite{Erickson:2001bq,DeDeo:2003te}, condensation of dark matter \cite{Bassett:2002fe} and the
Chaplygin gas, in terms of the matter power spectrum
\cite{Sandvik:2002jz,Beca:2003an} and combined full CMB and large scale
structure measurements \cite{Bean:2003fb,Amendola:2003bz}.   
Let us also emphasize that dark energy perturbations may not be consistently set to zero in perturbation theory  
\cite{Kunz:2006wc} even if $w= -1$. Indeed, it is unavoidable that dark energy perturbations are generated, even if set
to zero on some initial hypersurface, due to the presence of a non--vanishing gravitational potential. Therefore, the
expression (\ref{eqlind}) rigorously holds only in the physical 
limit in which the speed of sound is very close to unity
\mbox{(if $w\neq -1$)} so that dark energy perturbations are sufficiently suppressed.   

In this Letter we study the effect of dark energy perturbations on the 
growth index $\gamma$. Our main motivation is to understand if the introduction of the new degrees of freedom
introduced by dark energy perturbations imply changes in $\gamma$ large compared to the
forecasted errors $\Delta\gamma\simeq {\cal O}(0.04)$ (at 68\% confidence level). 
Following the common lore, see for instance \cite{Bean:2003fb}, and to simplify the analysis, 
 we will assume that the speed of sound
associated with the dark energy perturbations and the equation of state do
not change appreciably in the proper time range 
and that the dark energy perturbations
have no shear. This is a good approximation in linear perturbation theory for dark energy models
with a scalar field. Under these assumptions, we  provide an analytical formula for the growth index $\gamma$ as a function of the
speed of sound, the equation of state $w$, the dark matter abundance and the comoving scale.
As we will see, in the presence of dark energy perturbations, the growth index
differs from the corresponding value without dark energy perturbations by an 
amount which is comparable to the  realistic forecasted errors, especially for small speed of sound and $w$ significantly
different from $-1$. This opens up the possibility that the presence of dark energy perturbations may leave a significant imprint
on the growth function of dark matter perturbations. 

The Letter is organized as follows. In Section 2 we summarize our framework and provide the necessary equations for the
perturbations at the linear level. In Section 3 we discuss the growth index and in Section 4 we give our results and summarize of our work. 

\section{The basic equations} \noindent In this section we shortly
describe how to obtain the second order differential equations describing
the evolution of the coupled linear perturbations of dark matter and dark
energy in a spatially flat Friedmann-Lema\^itre-Robertson-Walker (FLRW)
background. We will closely follow \cite{Ma:1995ey} and \cite{Bean:2003fb}
and work in the synchronous gauge for convenience. With this choice the
perturbed metric in comoving coordinates reads \be \label{pertmetric} \rmd
s^2 = a^2(\tau)\left[-\rmd \tau^2 + \left(\delta_{ij}+h_{ij}\right)\rmd
x^i \rmd x^j\right]\;, \ee where $h_{ij}$ encodes the perturbation and can
be decomposed into a trace part $h\equiv h^i_i$ and a traceless one. The background
equations are simply \bea \label{b1} 3\ch^2 &=& 8 \pi G a^2
\bar{\rho}\;,\\ \label{b2} 2 \ch' &=& - \ch^2 \left(1+3 w \Omega_x
\right)\,, \eea where $G$ denotes Newton's constant,
$\bar{\rho}=\bar{\rho_c}+\bar{\rho_x}$ is the total energy density, the
comoving Hubble parameter is $ \ch \equiv a'/a $, primes denote
derivatives with respect to the comoving time $\tau$ and we define the
time varying relative dark energy density as
$\Omega_x=\bar{\rho_x}/\bar{\rho}\,$. The bars indicate homogeneous
background quantities and the subindexes `$_c$' and `$_x$' refer to dark
matter and dark energy respectively. We assume that the equation of state
of dark energy, $w$, is a constant and that the dark energy and the dark
matter do not interact. The divergence of the dark matter
velocity in its own rest frame is zero by definition and therefore in
Fourier space we have \be \label{restframe} \delta_c'+\frac{1}{2}h'=0\,,
\ee where \be \delta\rho_c\equiv{\bar\rho_c}\;\delta_c\;, \ee is the
energy density perturbation of dark matter.  The speed of sound of a fluid
can be defined as the ratio \cite{Bean:2003fb} \be \label{soundef}
c_s^2\equiv\frac{\delta P}{\delta \rho}\,, \ee where we have introduced
$\delta P$, the pressure perturbation of the fluid. It is important to
recall that the speed of sound defined in this way is a gauge dependent
quantity. However, the speed of sound is gauge invariant when measured in
the rest frame of the fluid. The
pressure perturbation of a dark energy component with constant equation of
state can be written in any reference frame in terms of its rest frame
speed of sound $\hat{c}_s$ as follows \be \label{pressper} \delta P_x =
\hat{c}_s^2 \delta \rho_x + 3 \ch
\left(1+w\right)\left(\hat{c}_s^2-w\right)\rho_x\frac{\theta_x}{k^2}\,,
\ee where $\theta_x$ is the dark energy velocity perturbation and $k$ the
inverse distance scale coming from the Fourier transformation. Then,
taking into account this expression and the relation \be h''+\ch h'=8\pi G a^2
\left(\delta T^0_0-\delta T^i_i\right)\,, \ee where $T^\mu_\nu$ is the
energy-momentum tensor, one can differentiate (\ref{restframe}) with
respect to $\tau$ and make use of the background evolution (\ref{b1}),
(\ref{b2}) to find the equation for the dark matter energy density
perturbation \cite{Bean:2003fb} \be \label{first} \dc''+\ch
\dc'-\frac{3}{2}\ch^2\Omega_c\dc=\frac{3}{2}\ch^2\Omega_x\left[\left(1+3\cshat\right)\dx+9\left(1+w\right)\ch\left(\cshat-w\right)\frac{\theta_x}{k^2}\right]\;.
\ee The time derivative of the dark energy density perturbation in the
dark matter rest frame is \cite{Bean:2003fb} \be \label{second}
\dx'=-(1+w)\left\{\left[k^2+9\left(\cshat-w\right)\ch^2\right]\frac{\theta_x}{k^2}-\dc'\right\}-3\ch(\cshat-w)\dx
\ee and the time derivative of the divergence of the dark energy velocity
perturbation in the case of no anisotropic stress perturbation is \be
\label{depvel}
\frac{\theta_x'}{k^2}=-\left(1-3\cshat\right)\ch\frac{\theta_x}{k^2}+\frac{\cshat}{1+w}\delta_x\;.
\ee Differentiating (\ref{second}) with respect to the comoving time and
combining (\ref{second}) and (\ref{depvel}) with the background equations
into the resulting expression one gets \bea \label{third} \nonumber
\dx''&+&\left[3\left(\cshat-w\right)\ch-\fac\right]\dx'\\ \nonumber
&+&\left\{\cshat
k^2-\frac{3}{2}\left(\cshat-w\right)\ch\left[\left(1+3w\Omega_x-6\cshat\right)\ch
+2 \fac\right]\right\} \dx\\&=& (1+w) \dc'' -(1+w) \fac \dc'\;, \eea where
\be \label{fourth} \fac\equiv
-9\left(1+3w\Omega_x\right)\frac{\cshat-w}{k^2+9\left(\cshat-w\right)\ch^2}\ch^3-(1-3\cshat)\ch\;.
\ee Equations (\ref{first}), (\ref{second}), (\ref{third}) and
(\ref{fourth}) allow us to describe the evolution of linear perturbations
of dark matter and dark energy as functions of time in a FLRW background.
Initial conditions are given at the redshift $z_{mr}=3200$, which
approximately corresponds to the time of matter-radiation equality. Since
we consider non--interacting fluids to describe the dark matter and dark
energy, their energy densities satisfy: 
\bea
\bar{\rho_c}'+3\ch\bar{\rho_c}&=&0\;,\\
\bar{\rho_x}'+3(1+w)\ch\bar{\rho_x}&=&0. \eea 
We choose adiabatic initial
conditions \be \delta_{x(mr)}=(1+w)\delta_{c(mr)}\;. \ee 
Furthermore, we assume zero initial time derivatives of the matter and
dark energy perturbations.  This is consistent with the fact
that at early times (both in the radiation and matter dominated
periods) the equations of the perturbations admit the solution \mbox{$\dx\propto(1+w)\dc\propto\tau^2$}
\cite{Bean:2003fb} as can be checked with (\ref{first}),
(\ref{second}), (\ref{third}) and (\ref{fourth}) and
and we have set the initial conformal time to zero. In fact we can even use
non--zero initial velocities and consider non--adiabatic initial conditions;
our results on the growth index are robust under these modifications. For
the background we consider the present (i.e. at $z_0=0, a_0\equiv1$) value
of the relative energy density of dark matter in the range $(0.25,0.30)$
and $\Omega_x^0=1-\Omega_c^0$. In our computations we do not include a
specific baryon component. We have checked that the effect of
adding baryons on the growth index can be at most as big as 0.2\%, which
is much smaller than the 8\% accuracy forecasted for the near future
experiments.

\section{The growth index} \noindent The growth of matter perturbations
has been studied neglecting the effect of dark energy perturbations through
the behaviour of the growth function \cite{Wang:1998gt} \be \label{growth}
g\equiv\frac{\dc}{a}\; \ee as a function of the natural logarithm of the
scale factor. It is possible to fit $g$ using a
simple parameterization that defines the growth index $\gamma$ and depends on the relative energy density of dark
matter  
\be \label{gammadef}
g(a)=g(a_i)\exp\int_{a_i}^a
\left(\Omega_c(\tilde{a})^\gamma-1\right)\frac{\rmd\tilde{a}}{\tilde{a}}\;.
\ee
The growth function depends on the scale $k$, the sound speed $\cshat$
and the equation of state $w$. This dependence is embedded in the
growth index $\gamma$ which therefore from now on has to be understood
as a function of these parameters. The growth factor $g$ can be normalized to unity at some $a_i>a_{(mr)}$
deep in the matter dominated epoch where $\delta_c\sim a$. The growth
index $\gamma$ is normally taken to be a (model--dependent) number whose best
fitting value for standard gravity and no dark energy perturbations is
around 0.55, see equation (\ref{eqlind}). This result is obtained from the
equation  
\be \label{justmatter}
\dc''+\ch\dc'-\frac{3}{2}\ch^2\Omega_c\dc=0\,, 
\ee with no dark energy perturbations, instead of the system of
second order differential equations that includes $\dx\,$. 

It is important to remark that it is not possible to reduce the system (\ref{first}),
(\ref{second}), (\ref{third}) and (\ref{fourth}) to (\ref{justmatter}) by
setting $\delta_x=0$ or with any particular choice of the parameters. Those
equations show that even if the dark energy perturbation is set to zero
initially it will be generated at later times. The effect of dark energy perturbations should be
included in the analysis of the growth history for consistency.
The growth of dark matter perturbations depends not only on $w$ (which already enters in (\ref{justmatter}) through $\Omega_c$ and $\ch$) but also on the other two parameters appearing explicitly in the differential equations that control the evolution of the perturbations, i.e. $k$ and $\cshat\,$. The reason for the dependence of the dark matter perturbations on the sound speed of dark energy is clear from the previous discussion and the definition (\ref{soundef}). In contrast to equation (\ref{justmatter}), the dependence on the comoving momentum 
now appears explicitly as an effect of a non--vanishing speed of sound. 

Given the numerical solution for the dark matter perturbation evolution, the definition (\ref{gammadef}) of the growth index can be used to compute $\gamma$ exactly:
\be
\label{eqgamma}
\gamma=\left(\ln \Omega_c\right)^{-1} \ln\left(\frac{a}{\dc}\frac{\rmd \dc}{\rmd a}\right)\;.
\ee
In the next section we will use this equation together with (\ref{first}),
(\ref{second}), (\ref{third}) and (\ref{fourth}) for obtaining our results. Obviously $\gamma$ will be a function of $a$ and it will depend on $k$, $\cshat$, $w$ and $\Omega_c^0$ as well.

In our analysis we consider $w$ in the reasonably broad range $(-1,-0.7)$. We choose not to allow the possibility that the equation of state of dark energy can be smaller than $-1$.  As for $k$, the values of interest  are the ones for which there is large scale structure data on the matter power spectrum \cite{Tegmark:2006az}. This goes approximately from $0.01 h\; {\rm Mpc}^{-1}$ to $0.2 h\; {\rm Mpc}^{-1}$,  including the nonlinear part of the spectrum which becomes so at roughly $0.09 h\; {\rm Mpc}^{-1}$. The scale that corresponds to the Hubble size today is $2.4\;10^{-4}\; {\rm Mpc}^{-1}$ and if we normalize it to $\ch_0=1\,$, the range of $k$ we will focus on (discarding the nonlinear part of the spectrum) is approximately $(30,270)$ in units of $\ch_0\,$. Notice that the lower $k$ value roughly gives the position of the baryon acoustic oscillation peak that can be used for constraining the growth index \cite{Sapone:2007jn}. Finally, regarding the sound speed of dark energy, we restrict $\cshat$ to be positive and smaller or equal than unity as 
currently  the bound  is very weak \cite{Weller:2003hw, Bean:2003fb,
Hannestad:2005ak, Xia:2007km, Mota:2007sz, TorresRodriguez:2008et}.

\begin{figure}
\begin{center}
\psfrag{z}[b][t][1.3][0]{$\gamma(z=1)$}   
\psfrag{w}[t][t][1.3][0]{$w$}
\includegraphics[width=0.9\linewidth]{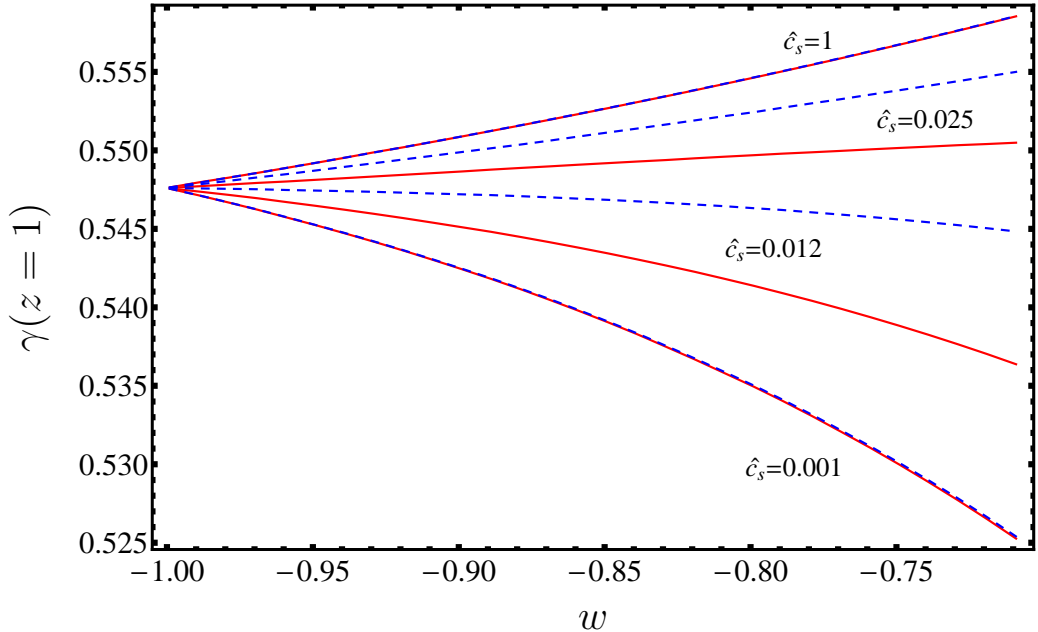}
\caption{\small{$\gamma(z=1)$ as a function of $w$ is shown for four values of $\hat{c}_s$. Red curves correspond to $k=0.050\, h {\rm Mpc}^{-1}$ and blue dashed ones to $k=0.078\, h {\rm Mpc}^{-1}$.}}
\label{functionwf}
\end{center}
\end{figure}

\section{Results and discussion}
\noindent
In this section we present a combination of numerical results and an analytical formula for the 
growth index $\gamma$ as a function of the relevant cosmological parameters. 

In Figure~\ref{functionwf} we plot the growth index at $z=1$ versus $w$ for several values of the speed of sound of dark energy and two different scales. Notice that the curves for the two different values of the comoving momenta coincide for $\hat{c}_s=1$ and in the limit of very small speed of sound. The figure indicates that the dark energy speed of sound and the scale determine whether $\gamma$ grows or decreases as a function $w$ at a given redshift. This is one of the reasons why having a more complete parameterization than (\ref{eqlind}) is important. Choosing another redshift would have the effect of an overall shift of the merging point at $w=-1$ together with modifications in the curvatures of the lines. 

\begin{figure}
\begin{center}
\psfrag{a}[t][t][1.3][0]{$w$}   
\psfrag{b}[b][t][1.3][0]{$\left(\gamma(z=1)/\gamma_{ap}(z=1)-1\right)\times100$}              
\includegraphics[width=0.9\linewidth]{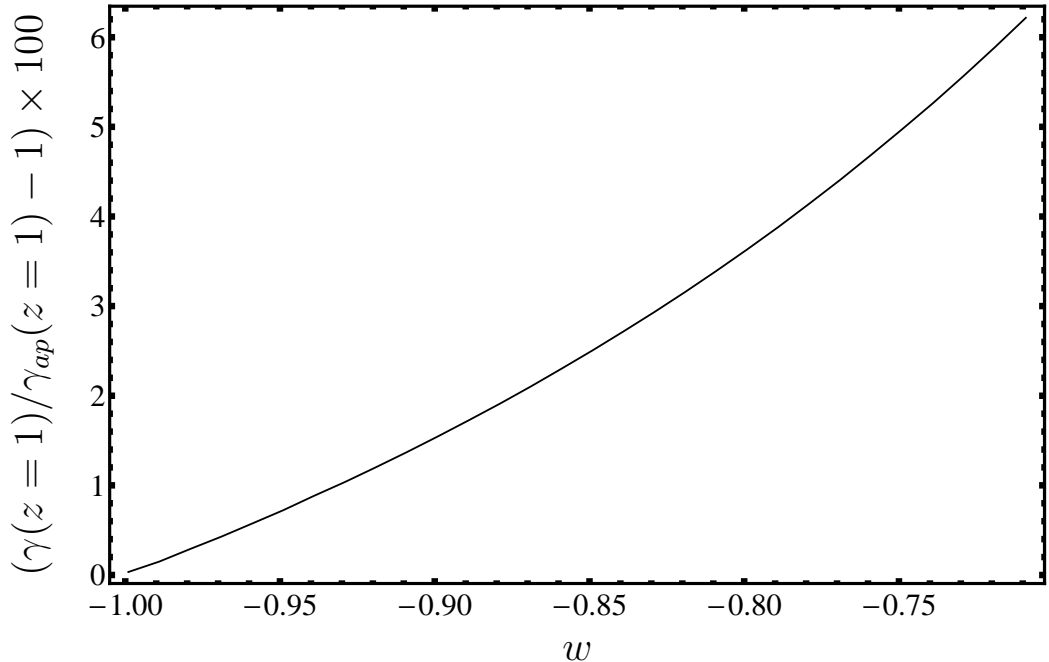}
\caption{\small{Relative error as a function of $w$ between the exact numerical result for $\gamma(z=1)$ with very small dark energy speed of sound and the approximation $\gamma_{ap}$ at the same redshift based on Equations (\ref{aa}) and (\ref{first}) with zero $\theta_x$. The figure has been done for $\cshat=10^{-6}$, $\Omega_c^0=0.30$ and $k=0.050\, h {\rm Mpc}^{-1}\,$.}}
\label{diff}
\end{center}
\end{figure}

To gain some insight on the change of the value of $\gamma$ from $\cshat=1$ to $\cshat\ll 1$, we observe that, 
in the limit $\cshat\simeq 0$ and 
from Equation (\ref{depvel}),  the dark energy velocity 
perturbation promptly decays in time. One is left with the following
solution for $\delta_x$
\begin{equation}
\label{aa}
\delta_x(a)=\delta_{x(mr)}\left(\frac{a}{a_{(mr)}}\right)^{3w}+
(1+w)a^{3w}\int \, 
\tilde{a}^{-3w-1}\,\dot{\delta}_c\left(\cshat=1\right)\,{\rm d}\tilde{a}\,,
\end{equation}
where the dot  stands for differentiation with respect to $\ln a$. As a first approximation, we can solve Equation (\ref{aa}) plugging in the dark matter perturbation $\delta_c\left(\cshat=1\right)$ obtained taking $\cshat=1$, which for this purpose corresponds to the case in which no dark energy
perturbations are present. From Equation (\ref{first}), it is clear  that the dark energy perturbations provide an extra source for the
dark matter pertrubation growth. We then solve numerically Equation (\ref{first}) with this new known source and $\theta_x=0\,$.
The difference between the true value of $\gamma$ and the one obtained with such an approximation is plotted
in Figure~\ref{diff}. 

\begin{figure}
\begin{center}
\psfrag{z}[b][t][1.3][0]{$\gamma(z=1)$}   
\psfrag{w}[t][t][1.3][0]{$\log_{10}\hat{c}_s$}
\psfrag{A}[c][c][0.8][0]{$w=-0.99$}    
\psfrag{B}[c][c][0.8][0]{$w=-0.9$}  
\psfrag{C}[c][c][0.8][0]{$w=-0.8$}  
\psfrag{D}[c][c][0.8][0]{$w=-0.7$}  
\includegraphics[width=0.9\linewidth]{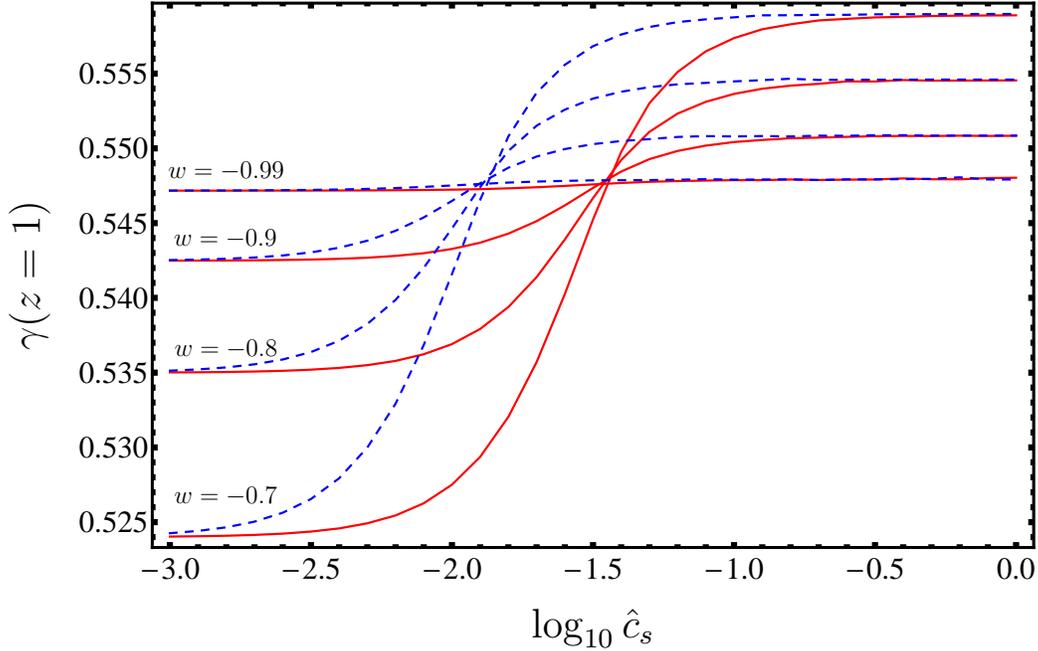}
\caption{\small{$\gamma(z=1)$ as a function of $\log_{10}\hat{c}_s$ is shown for four values of $w$. Red curves correspond to $k=0.03\, h {\rm Mpc}^{-1}$ and blue dashed ones to $k=0.08\, h {\rm Mpc}^{-1}$.}}
\label{functioncs}
\end{center}
\end{figure}

In Figure~\ref{functioncs} we show the growth index at $z=1$ versus $\log_{10}\hat{c}_s$ for different values of the equation of state of dark energy and two scales $k$. From this plot it is clear that the effect of changing the scale is an overall shift along the $\log_{10}\hat{c}_s$ axis. Notice that the intersecting points for the two sets of lines have the same value of the growth index, $\gamma\simeq0.547$, which corresponds to the merging point in Figure~\ref{functionwf}.

The redshift dependence of the growth index has already been studied without taking into account dark energy perturbations \cite{Polarski:2007rr} concluding that $\rmd \gamma /\rmd z\sim-0.02$ at $z=0\,$; being this value nearly independent of $z$ for a given $\Omega_c^0\,$. However, including dark energy perturbations, we find that it is actually the derivative of $\gamma$ with respect to the scale factor $a$ which is constant. Therefore the redshift dependence of the growth index can be better modeled with a $1/z$ term plus a constant term. We will later see that the growth index actually has an almost constant slope as a function of the scale factor when dark energy perturbations are taken into account.

Our next step is to obtain an analytical parameterization of the growth index as a function of the cosmological parameters. We  
start with the following generic ansatz:
\be
\label{linans}
\gamma\left(\Omega_c^0,\cshatn,k,w,a\right)=\gamma_{eq}\left(\Omega_c^0,\cshatn,k,w\right)+\zeta\left(\Omega_c^0,\cshatn,k,w\right)\left[a-a_{eq}\left(\Omega_c^0,w\right)\right]\;,
\ee
where $a_{eq}$ is the value of the scale factor at which ``dark equality'' ($\Omega_c=\Omega_x=1/2$) takes place:
\be
\label{equality}
a_{eq}=\left(\frac{1}{\Omega_c^0}-1\right)^{\frac{1}{3w}}\;.
\ee
We want to fit the growth index for $a$ in the interval $[a_{eq},1]$ which approximately corresponds to a redshift $z\in[0,0.55]$ for the ranges of the equation of state of dark energy and its relative energy density that we consider. Ideally one would wish to be able to use (\ref{eqgamma}) and the equations for the perturbations to infer completely the analytical dependence of $\gamma_{eq}\left(\Omega_c^0,\cshatn,k,w\right)$ and $\zeta\left(\Omega_c^0,\cshatn,k,w\right)$ in their variables. This turns out to be difficult and we find it efficient to make a numerical fit directly. The generic form (\ref{linans}), which can be viewed as a first order Taylor expansion in the scale factor, is motivated by the nearly zero variation of $\rmd \gamma /\rmd a$. The choice of $a_{eq}$ as the point around which we make the expansion is a convenient one,  but the fit could in principle be done taking a model independent value of $a$ as the fiducial point. We use the same ansatz to fit $\gamma_{eq}$ and $\gamma_0\,$, which is the growth index at $a_0=1$, and doing so we directly obtain the slope $\zeta\,$ from (\ref{linans}):
\be\label{slope}
\zeta\left(\Omega_c^0,\cshatn,k,w\right)=\frac{\gamma_0-\gamma_{eq}}{1-a_{eq}}\;.
\ee
In particular, we assume the following parameterization for $\gamma_{eq}$ and $\gamma_0$:
\be\label{ordinate}
\gamma_{j}\left(\cshatn,k,w\right)=h_{j}(w)\tanh\left[\left(\log_{10} \cshatn - g_{j}(k)\right)\frac{r_{j}(w)}{h_{j}(w)}\right]+f_{j}(w)\;,\, j=\{eq,0\}\,.
\ee
Notice that we have taken $\gamma_{eq}$ and $\gamma_0$ to be independent of $\Omega_c^0\,$ and we incorporate this assumption in our notation, so we will refer to $\gamma_{j}\left(\cshatn,k,w\right)$ from now on. The functions $f_{j}(w)\,$, $g_{j}(k)\,$, $h_{j}(w)\,$ and $r_{j}(w)\,$ are polynomials in their variables. It turns out that the fit obtained with this procedure can be importantly improved with the addition of a polynomial correction to $\zeta$ that depends on $\Omega_c^0$, so finally:
\bea
\nonumber
\gamma\left(\Omega_c^0,\cshatn,k,w,a\right)&=&\gamma_{eq}\left(\cshatn,k,w\right)\\ \label{gammacorr} &&+\left[\zeta\left(\Omega_c^0,\cshatn,k,w\right)+\eta\left(\Omega_c^0\right)\right]\left[a-a_{eq}\left(\Omega_c^0,w\right)\right]\,.
\eea

\begin{figure}
\begin{center}
\psfrag{z}[t][t][1.3][0]{$\log_{10}\hat{c}_s$}   
\psfrag{g}[b][t][1.3][0]{$\gamma_{eq}\left(\cshatn,k,w\right)$} 
\psfrag{A}[c][c][0.8][0]{A}    
\psfrag{B}[c][c][0.8][0]{B}  
\psfrag{C}[c][c][0.8][0]{C}  
\psfrag{D}[c][c][0.8][0]{D}      
\includegraphics[width=0.9\linewidth]{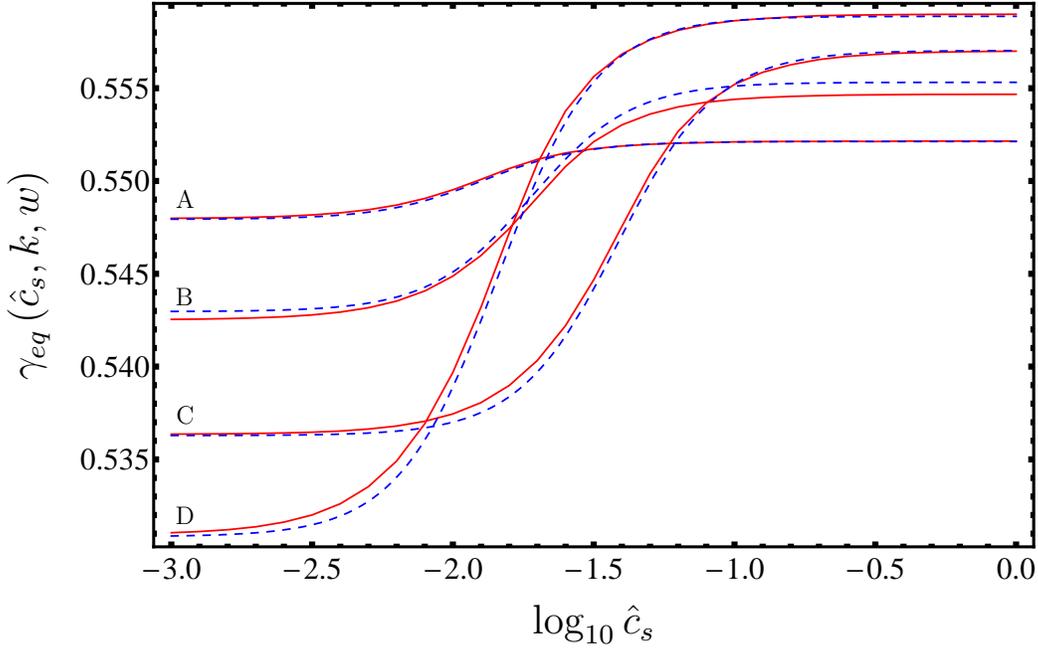}
\caption{\small{$\gamma_{eq}\left(\cshatn,k,w\right)$ versus $\log_{10}\hat{c}_s$ for different combinations of the pair $\{k,w\}
\,$: \newline ${\rm A}=\{0.08\, h\, {\rm Mpc^{-1}},-0.95\}\,$, ${\rm B}=\{0.02\, h\, {\rm Mpc^{-1}},-0.7\}\,$, ${\rm C}=\{0.04\, h\, {\rm Mpc^{-1}},-0.87\}\,$ and ${\rm D}=\{0.06\, h\, {\rm Mpc^{-1}},-0.75\}$. Red lines are the exact numerical result and blue dashed ones the corresponding fits.}}
\label{figeq}
\end{center}
\end{figure}

The set of equations (\ref{slope}), (\ref{ordinate}) and (\ref{gammacorr}) constitute the full fitting formula for the growth index. The resulting nine polynomials through which the fit can be expressed are the following:
{\small
\bea
f_{eq}(w)&=&4.498\cdot10^{-1}- 2.176\cdot10^{-1}\, w - 1.041\cdot10^{-1}\, w^2 + 5.287\cdot10^{-2}\,w^3\nonumber\\&+&4.030\cdot10^{-2}\,w^4\,,\\
f_{0}(w)&=&4.264\cdot10^{-1}- 3.217\cdot10^{-1}\, w - 2.581\cdot10^{-1}\, w^2- 5.512\cdot10^{-2}\,w^3\nonumber\\&+&1.054\cdot10^{-2}\,w^4\,,
\eea
\bea
g_{eq}(k)&=&-5.879\cdot10^{-1} - 2.296\cdot10^{-2}\, k + 2.125\cdot10^{-4}\, k^2 - 1.177\cdot10^{-6}\, k^3\nonumber\\  &+& 
3.357\cdot10^{-10}\, k^4-3.801\cdot10^{-12}k^5\,,\\
g_{0}(k)&=&-6.401\cdot10^{-1} - 2.291\cdot10^{-2}\, k + 2.119\cdot10^{-4}\, k^2 - 1.173\cdot10^{-6}\, k^3\nonumber\\ &+& 
 3.344\cdot10^{-10}\, k^4-3.787\cdot10^{-12}k^5\,,
\eea
\bea
h_{eq}(w)&=&1.759\cdot10^{-1} + 4.066\cdot10^{-1}\, w + 3.254\cdot10^{-1}\, w^2+9.470\cdot10^{-2}\, w^3\,,\\
h_{eq}(w)&=&2.008\cdot10^{-1} + 4.644\cdot10^{-1}\, w + 3.713\cdot10^{-1}\, w^2+1.076\cdot10^{-1}\, w^3\,,
\eea
\bea
r_{eq}(w)&=& 5.158\cdot10^{-1} + 1.203 w + 9.697\cdot10^{-1}\, w^2+2.827\cdot10^{-1}\, w^3\,,\\
r_{0}(w)&=& 6.093\cdot10^{-1} + 1.435 w +  1.1668 w^2+3.412\cdot10^{-1}\, w^3\,,
\eea
\be
\eta(\Omega_c^0)= 8.037\cdot10^{-3} + 4.676\cdot10^{-2}\,\Omega_c^0  - 2.829\cdot10^{-1} \left(\Omega_c^0\right)^2.
\ee
}
The truncation of the coefficients above has been done in such a way that the figures in the Letter can be reproduced and that the maximum relative error between the numerical value of $\gamma$ and the fitting formula does not exceed 0.2\% for any combination of the parameters. In fact, this error turns out to be much smaller for generic choices of the parameters.

\begin{figure}
\begin{center}
\psfrag{a}[t][t][1.2][0]{$a$}   
\psfrag{b}[b][t][1.2][0]{$\gamma\left(\Omega_c^0,\cshatn,k,w,a\right)$}  
\psfrag{r1}[c][c][0.8][0]{$w=-0.70$}      
\psfrag{r2}[c][c][0.8][0]{$w=-0.75$}      
\psfrag{r3}[c][c][0.8][0]{$w=-0.80$}      
\psfrag{r4}[c][c][0.8][0]{$w=-0.90$}      
\psfrag{r5}[c][c][0.8][0]{$w=-0.99$}            
\includegraphics[width=0.9\linewidth]{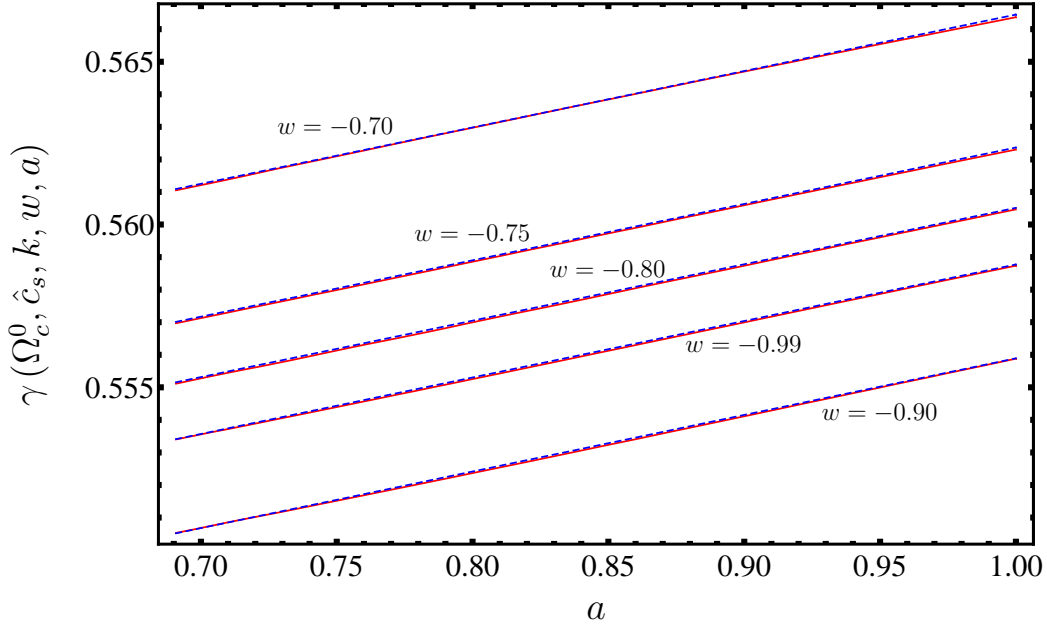}
\caption{\small{$\gamma\left(\Omega_c^0,\cshatn,k,w,a\right)$ versus $a$ for $k=0.033\, h\, {\rm Mpc^{-1}}\,$, $\Omega_c^0=0.27\,$ and $\cshat=0.01\,$. Different values of $w$ are chosen as shown in the figure. The red lines are the numerical results from the differential equations and the blue dashed ones are the fits to them.}}
\label{figuraw}
\end{center}
\end{figure}

\begin{figure}
\begin{center}
\psfrag{a}[t][t][1.3][0]{$a$}   
\psfrag{b}[b][t][1.3][0]{$\gamma\left(\Omega_c^0,\cshatn,k,w,a\right)$}              
\includegraphics[width=0.9\linewidth]{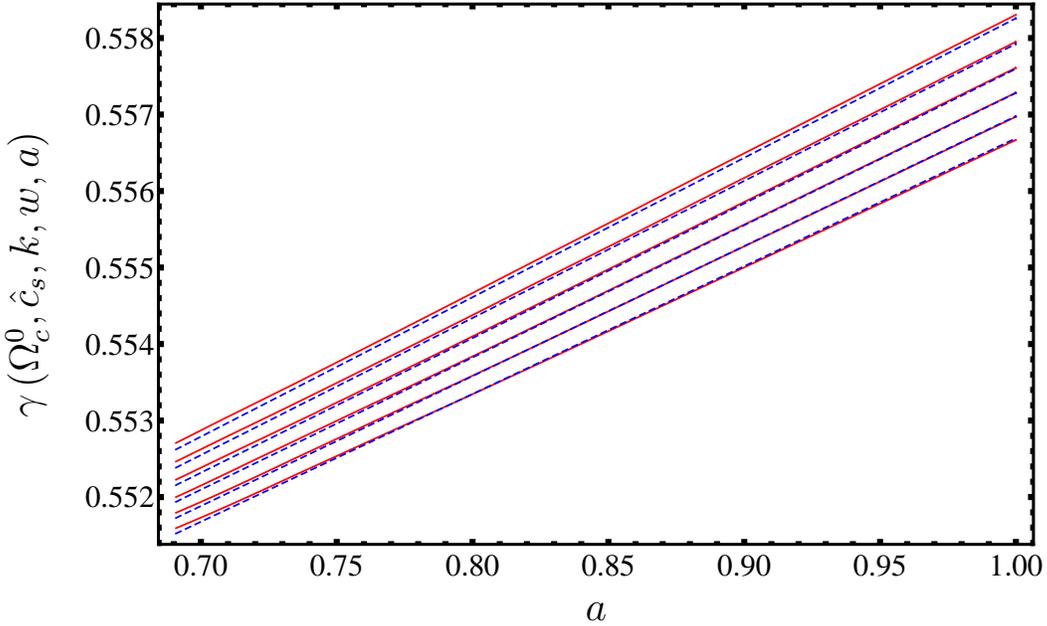}
\caption{\small{$\gamma\left(\Omega_c^0,\cshatn,k,w,a\right)$ versus $a$ for $k=0.03\, h\, {\rm Mpc^{-1}}\,$, $w=-0.92$ and $\cshat=0.0036\,$. The value of $\Omega_c^0\,$ runs between 0.25 and 0.30 in steps of 0.01 from top to bottom of the figure. The red lines are the numerical results from the differential equations and the blue dashed ones are the fits to them.}}
\label{figuraOmega}
\end{center}
\end{figure}

\begin{figure}
\begin{center}
\psfrag{a}[t][t][1.3][0]{$a$}   
\psfrag{b}[b][t][1.3][0]{$\gamma\left(\Omega_c^0,\cshatn,k,w,a\right)$}              
\includegraphics[width=0.9\linewidth]{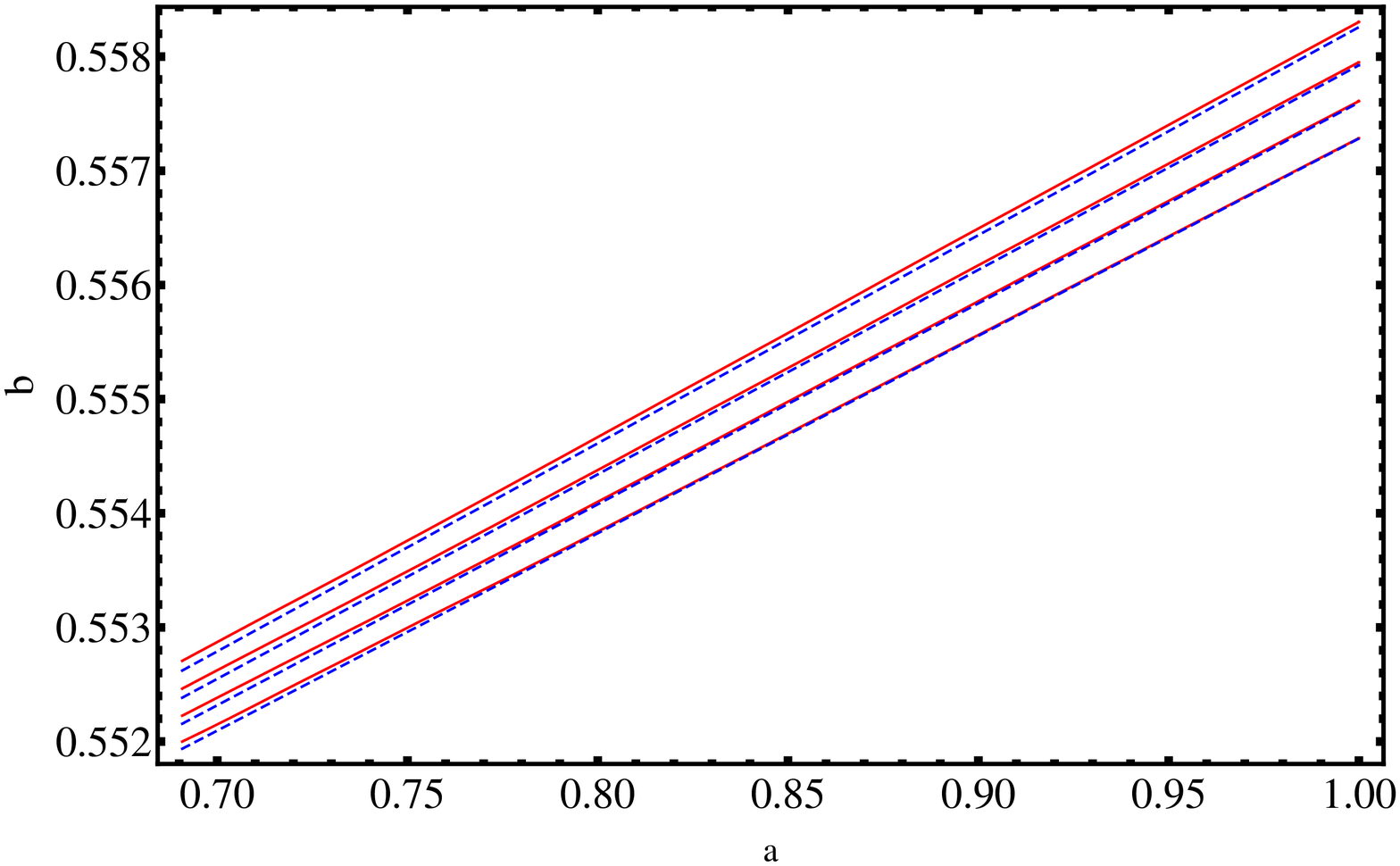}
\caption{\small{$\gamma\left(\Omega_c^0,\cshatn,k,w,a\right)$ versus $a$ for $w=-0.80\,$, $\cshat=0.01$ and $\Omega_c^0=0.27\,$. The scale $k$ in units of $h\, {\rm Mpc^{-1}}$ takes the values $\{0.023, 0.027, 0.037, 0.067 \}\,$ from bottom to top of the figure. The red lines are the numerical results from the differential equations and the blue dashed ones are the fits to them.}}
\label{figurak}
\end{center}
\end{figure}

In Figure~\ref{figeq} we show $\gamma_{eq}\left(\cshatn,k,w\right)$ versus the decimal logarithm of $\cshatn$ for several combinations of $k$ and $w$. The red curves represent the exact numerical growth index and the blue dashed lines are the corresponding fits. In Figures~\ref{figuraw}, \ref{figuraOmega} and \ref{figurak} we show $\gamma\left(\Omega_c^0,\cshatn,k,w,a\right)$ versus the scale factor for several values of $w$, $\Omega_c^0$ and $k$ respectively, as explained in the captions. The other parameters are kept fixed. The colour code, as in Figure~\ref{figeq}, is that the red curves represent the exact numerical growth index and the blue dashed lines are the corresponding fits. These figures are meant to illustrate the goodness of fit for several choices of the parameters. 

Equations (25)-(36) offer an analytical expression for the growth index in terms of the relevant cosmological parameters
in the case in which dark energy perturbations are present. 
The case without dark energy perturbations is reproduced
by assuming $\cshat=1\,$. 
The analytical parameterization
fits the numerical results  in the assumed range of parameters to a precision of 0.2\% (in the worst cases) or better for the growth index.
Our findings show that
$\gamma$ can vary from 0.55 by an amount $\Delta\gamma$ as large as $\sim0.03\,$. We have checked that this result holds for any redshift between $z_{eq}$ (at the time of dark equality) and $z=1$. This difference 
is of the same order of magnitude of  the 68\% c.l. forecasted error band. The predicted value of $\gamma$ may differ by this amount 
from the value without dark energy perturbations 
if the speed of sound is tiny and if the equation 
of state substantially deviates from $-1$. This opens up the
possibility that a detailed future measurement of the growth 
factor might help in revealing the presence of dark energy perturbations. 
Finally, let us reiterate that our results have been obtained 
under the assumption that $\cshat$ and $w$ 
do not evolve in time, at least for mild values of redshift. Furthermore, we
have assumed that  the dark energy perturbations
have no shear.

\section*{Acknowledgments}

This work has received financial support from the Spanish Ministry of Education and Science through
the research project FPA2004-02015; by the Comunidad de Madrid through project
P-ESP-00346; by a Marie Curie Fellowship of the European
Community under contract MEST-CT-2005-020238-EUROTHEPY; by the
European Commission under
contracts MRTN-CT-2004-503369 and MRTN-CT-2006-035863 (Marie Curie
Research and Training Network ``UniverseNet'') and by the Comunidad de Madrid and the European
Social Fund through a FPI contract.\\
Guillermo Ballesteros thanks the hospitality of the Theory Division at CERN. 

\bibliography{pertrefs}

\begin{thebibliography}{10}

\bibitem{Astier:2005qq}
P.~Astier {\em et~al.}, ``{The Supernova Legacy Survey: Measurement of
  $\Omega_M$, $\Omega_{\Lambda}$ and $w$ from the First Year Data Set},'' {\em
  Astron. Astrophys.}, vol.~447, pp.~31--48, 2006, astro-ph/0510447.

\bibitem{ArmendarizPicon:2000dh}
C.~Armendariz-Picon, V.~F. Mukhanov, and P.~J. Steinhardt, ``{A dynamical
  solution to the problem of a small cosmological constant and late-time cosmic
  acceleration},'' {\em Phys. Rev. Lett.}, vol.~85, pp.~4438--4441, 2000,
  astro-ph/0004134.

\bibitem{Copeland:2006wr}
E.~J. Copeland, M.~Sami, and S.~Tsujikawa, ``{Dynamics of dark energy},'' {\em
  Int. J. Mod. Phys.}, vol.~D15, pp.~1753--1936, 2006, hep-th/0603057.

\bibitem{Carroll:2003wy}
S.~M. Carroll, V.~Duvvuri, M.~Trodden, and M.~S. Turner, ``{Is cosmic speed-up
  due to new gravitational physics?},'' {\em Phys. Rev.}, vol.~D70, p.~043528,
  2004, astro-ph/0306438.

\bibitem{Dvali:2000hr}
G.~R. Dvali, G.~Gabadadze, and M.~Porrati, ``{4D gravity on a brane in 5D
  Minkowski space},'' {\em Phys. Lett.}, vol.~B485, pp.~208--214, 2000,
  hep-th/0005016.

\bibitem{Knox:2005rg}
L.~Knox, Y.-S. Song, and J.~A. Tyson, ``{Two windows on acceleration and
  gravitation: Dark energy or new gravity?},'' 2005, astro-ph/0503644.

\bibitem{Koyama:2005kd}
K.~Koyama and R.~Maartens, ``{Structure formation in the DGP cosmological
  model},'' {\em JCAP}, vol.~0601, p.~016, 2006, astro-ph/0511634.

\bibitem{Koyama:2006ef}
K.~Koyama, ``{Structure formation in modified gravity models alternative to
  dark energy},'' {\em JCAP}, vol.~0603, p.~017, 2006, astro-ph/0601220.

\bibitem{Kunz:2006ca}
M.~Kunz and D.~Sapone, ``{Dark energy versus modified gravity},'' {\em Phys.
  Rev. Lett.}, vol.~98, p.~121301, 2007, astro-ph/0612452.

\bibitem{Uzan:2006mf}
J.-P. Uzan, ``{The acceleration of the universe and the physics behind it},''
  {\em Gen. Rel. Grav.}, vol.~39, pp.~307--342, 2007, astro-ph/0605313.

\bibitem{Carroll:2006jn}
S.~M. Carroll, I.~Sawicki, A.~Silvestri, and M.~Trodden, ``{Modified-Source
  Gravity and Cosmological Structure Formation},'' {\em New J. Phys.}, vol.~8,
  p.~323, 2006, astro-ph/0607458.

\bibitem{Bertschinger:2006aw}
E.~Bertschinger, ``{On the Growth of Perturbations as a Test of Dark Energy},''
  {\em Astrophys. J.}, vol.~648, pp.~797--806, 2006, astro-ph/0604485.

\bibitem{Linder:2007hg}
E.~V. Linder and R.~N. Cahn, ``{Parameterized Beyond-Einstein Growth},'' {\em
  Astropart. Phys.}, vol.~28, pp.~481--488, 2007, astro-ph/0701317.

\bibitem{Tsujikawa:2007gd}
S.~Tsujikawa, ``{Matter density perturbations and effective gravitational
  constant in modified gravity models of dark energy},'' {\em Phys. Rev.},
  vol.~D76, p.~023514, 2007, 0705.1032.

\bibitem{Acquaviva:2007mm}
V.~Acquaviva and L.~Verde, ``{Observational signatures of Jordan-Brans-Dicke
  theories of gravity},'' {\em JCAP}, vol.~0712, p.~001, 2007, 0709.0082.

\bibitem{Jain:2007yk}
B.~Jain and P.~Zhang, ``{Observational Tests of Modified Gravity},'' 2007,
  0709.2375.

\bibitem{Wei:2008vw}
H.~Wei and S.~N. Zhang, ``{How to Distinguish Dark Energy and Modified
  Gravity?},'' 2008, 0803.3292.

\bibitem{Wei:2008ig}
H.~Wei, ``{Growth Index of DGP Model and Current Growth Rate Data},'' {\em
  Phys. Lett.}, vol.~B664, pp.~1--6, 2008, 0802.4122.

\bibitem{Zhang:2008kx}
H.~Zhang, H.~Yu, H.~Noh, and Z.-H. Zhu, ``{Probing the nature of cosmic
  acceleration},'' 2008, 0806.4082.

\bibitem{Bertschinger:2008zb}
E.~Bertschinger and P.~Zukin, ``{Distinguishing Modified Gravity from Dark
  Energy},'' 2008, 0801.2431.

\bibitem{Linder:2005in}
E.~V. Linder, ``{Cosmic growth history and expansion history},'' {\em Phys.
  Rev.}, vol.~D72, p.~043529, 2005, astro-ph/0507263.

\bibitem{Huterer:2006mva}
D.~Huterer and E.~V. Linder, ``{Separating dark physics from physical darkness:
  Minimalist modified gravity vs. dark energy},'' {\em Phys. Rev.}, vol.~D75,
  p.~023519, 2007, astro-ph/0608681.

\bibitem{Amendola:2007rr}
L.~Amendola, M.~Kunz, and D.~Sapone, ``{Measuring the dark side (with weak
  lensing)},'' {\em JCAP}, vol.~0804, p.~013, 2008, 0704.2421.

\bibitem{Sapone:2007jn}
D.~Sapone and L.~Amendola, ``{Constraining the growth factor with baryon
  oscillations},'' 2007, 0709.2792.

\bibitem{Nesseris:2007pa}
S.~Nesseris and L.~Perivolaropoulos, ``{Testing LCDM with the Growth Function
  $\delta(a)$: Current Constraints},'' {\em Phys. Rev.}, vol.~D77, p.~023504,
  2008, 0710.1092.

\bibitem{Dore:2007jh}
O.~Dore {\em et~al.}, ``{Testing Gravity with the CFHTLS-Wide Cosmic Shear
  Survey and SDSS LRGs},'' 2007, 0712.1599.

\bibitem{Mantz:2007qh}
A.~Mantz, S.~W. Allen, H.~Ebeling, and D.~Rapetti, ``{New constraints on dark
  energy from the observed growth of the most X-ray luminous galaxy
  clusters},'' 2007, 0709.4294.

\bibitem{DiPorto:2007ym}
C.~Di~Porto and L.~Amendola, ``{Observational constraints on the linear
  fluctuation growth rate},'' {\em Phys. Rev.}, vol.~D77, p.~083508, 2008,
  0707.2686.

\bibitem{Yamamoto:2007gd}
K.~Yamamoto, D.~Parkinson, T.~Hamana, R.~C. Nichol, and Y.~Suto, ``{Optimizing
  future imaging survey of galaxies to confront dark energy and modified
  gravity models},'' {\em Phys. Rev.}, vol.~D76, p.~023504, 2007, 0704.2949.

\bibitem{Heavens:2007ka}
A.~F. Heavens, T.~D. Kitching, and L.~Verde, ``{On model selection forecasting,
  dark energy and modified gravity},'' 2007, astro-ph/0703191.

\bibitem{Acquaviva:2008qp}
V.~Acquaviva, A.~Hajian, D.~N. Spergel, and S.~Das, ``{Next Generation Redshift
  Surveys and the Origin of Cosmic Acceleration},'' 2008, 0803.2236.

\bibitem{Refregier:2008js}
A.~Refregier and t.~D. collaboration, ``{The Dark UNiverse Explorer (DUNE):
  Proposal to ESA's Cosmic Vision},'' 2008, 0802.2522.

\bibitem{Gibbons:2002md}
G.~W. Gibbons, ``{Cosmological evolution of the rolling tachyon},'' {\em Phys.
  Lett.}, vol.~B537, pp.~1--4, 2002, hep-th/0204008.

\bibitem{Erickson:2001bq}
J.~K. Erickson, R.~R. Caldwell, P.~J. Steinhardt, C.~Armendariz-Picon, and
  V.~F. Mukhanov, ``{Measuring the speed of sound of quintessence},'' {\em
  Phys. Rev. Lett.}, vol.~88, p.~121301, 2002, astro-ph/0112438.

\bibitem{DeDeo:2003te}
S.~DeDeo, R.~R. Caldwell, and P.~J. Steinhardt, ``{Effects of the sound speed
  of quintessence on the microwave background and large scale structure},''
  {\em Phys. Rev.}, vol.~D67, p.~103509, 2003, astro-ph/0301284.

\bibitem{Bassett:2002fe}
B.~A. Bassett, M.~Kunz, D.~Parkinson, and C.~Ungarelli, ``{Condensate cosmology
  - Dark energy from dark matter},'' {\em Phys. Rev.}, vol.~D68, p.~043504,
  2003, astro-ph/0211303.

\bibitem{Sandvik:2002jz}
H.~Sandvik, M.~Tegmark, M.~Zaldarriaga, and I.~Waga, ``{The end of unified dark
  matter?},'' {\em Phys. Rev.}, vol.~D69, p.~123524, 2004, astro-ph/0212114.

\bibitem{Beca:2003an}
L.~M.~G. Beca, P.~P. Avelino, J.~P.~M. de~Carvalho, and C.~J. A.~P. Martins,
  ``{The Role of Baryons in Unified Dark Matter Models},'' {\em Phys. Rev.},
  vol.~D67, p.~101301, 2003, astro-ph/0303564.

\bibitem{Bean:2003fb}
R.~Bean and O.~Dor\'e, ``{Probing dark energy perturbations: the dark energy
  equation of state and speed of sound as measured by WMAP},'' {\em Phys.
  Rev.}, vol.~D69, p.~083503, 2004, astro-ph/0307100.

\bibitem{Amendola:2003bz}
L.~Amendola, F.~Finelli, C.~Burigana, and D.~Carturan, ``{WMAP and the
  Generalized Chaplygin Gas},'' {\em JCAP}, vol.~0307, p.~005, 2003,
  astro-ph/0304325.

\bibitem{Kunz:2006wc}
M.~Kunz and D.~Sapone, ``{Crossing the phantom divide},'' {\em Phys. Rev.},
  vol.~D74, p.~123503, 2006, astro-ph/0609040.

\bibitem{Ma:1995ey}
C.-P. Ma and E.~Bertschinger, ``{Cosmological perturbation theory in the
  synchronous and conformal Newtonian gauges},'' {\em Astrophys. J.}, vol.~455,
  pp.~7--25, 1995, astro-ph/9506072.

\bibitem{Wang:1998gt}
L.-M. Wang and P.~J. Steinhardt, ``{Cluster Abundance Constraints on
  Quintessence Models},'' {\em Astrophys. J.}, vol.~508, pp.~483--490, 1998,
  astro-ph/9804015.

\bibitem{Tegmark:2006az}
M.~Tegmark {\em et~al.}, ``{Cosmological Constraints from the SDSS Luminous Red
  Galaxies},'' {\em Phys. Rev.}, vol.~D74, p.~123507, 2006, astro-ph/0608632.

\bibitem{Weller:2003hw}
J.~Weller and A.~M. Lewis, ``{Large Scale Cosmic Microwave Background
  Anisotropies and Dark Energy},'' {\em Mon. Not. Roy. Astron. Soc.}, vol.~346,
  pp.~987--993, 2003, astro-ph/0307104.

\bibitem{Hannestad:2005ak}
S.~Hannestad, ``{Constraints on the sound speed of dark energy},'' {\em Phys.
  Rev.}, vol.~D71, p.~103519, 2005, astro-ph/0504017.

\bibitem{Xia:2007km}
J.-Q. Xia, Y.-F. Cai, T.-T. Qiu, G.-B. Zhao, and X.~Zhang, ``{Constraints on
  the sound speed of dynamical dark energy},'' 2007, astro-ph/0703202.

\bibitem{Mota:2007sz}
D.~F. Mota, J.~R. Kristiansen, T.~Koivisto, and N.~E. Groeneboom,
  ``{Constraining Dark Energy Anisotropic Stress},'' 2007, 0708.0830.

\bibitem{TorresRodriguez:2008et}
A.~Torres-Rodriguez, C.~M. Cress, and K.~Moodley, ``{Covariance of dark energy
  parameters and sound speed constraints from large HI surveys},'' 2008,
  0804.2344.

\bibitem{Polarski:2007rr}
D.~Polarski and R.~Gannouji, ``{On the growth of linear perturbations},'' {\em
  Phys. Lett.}, vol.~B660, pp.~439--443, 2008, 0710.1510.

\end{thebibliography}

\end{document}